\documentclass[english,aps,superscriptaddress,preprintnumbers]{revtex4}
\usepackage{color}
\usepackage{verbatim}
\usepackage{amsthm}
\usepackage{amsmath}
\usepackage{graphicx}
\usepackage{esint}

\makeatletter
\@ifundefined{textcolor}{}
{%
 \definecolor{BLACK}{gray}{0}
 \definecolor{WHITE}{gray}{1}
 \definecolor{RED}{rgb}{1,0,0}
 \definecolor{GREEN}{rgb}{0,1,0}
 \definecolor{BLUE}{rgb}{0,0,1}
 \definecolor{CYAN}{cmyk}{1,0,0,0}
 \definecolor{MAGENTA}{cmyk}{0,1,0,0}
 \definecolor{YELLOW}{cmyk}{0,0,1,0}
}
\numberwithin{equation}{section}
\numberwithin{figure}{section}

\makeatother

\usepackage{babel}
\begin{document}

\title{Wigner functions for fermions in strong magnetic fields}

\author{Xin-li Sheng}
\affiliation{Interdisciplinary Center for Theoretical Study and Department of
Modern Physics, University of Science and Technology of China, Hefei,
Anhui 230026, China}

\author{Dirk H.\ Rischke}
\affiliation{Interdisciplinary Center for Theoretical Study and Department of
Modern Physics, University of Science and Technology of China, Hefei,
Anhui 230026, China}
\affiliation{Institute for Theoretical Physics, Goethe University,
Max-von-Laue-Str.\ 1, D-60438 Frankfurt am Main, Germany}

\author{David Vasak}
\affiliation{Frankfurt Institute for Advanced Studies (FIAS),  Ruth-Moufang-Str.\ 1, 
D-60438 Frankfurt am Main, Germany}

\author{Qun Wang}
\affiliation{Interdisciplinary Center for Theoretical Study and Department of
Modern Physics, University of Science and Technology of China, Hefei,
Anhui 230026, China}

\begin{abstract}
We compute the covariant Wigner function for spin-1/2 fermions in
an arbitrarily strong magnetic field by exactly solving the Dirac
equation at non-zero fermion-number and chiral-charge densities. 
The Landau energy levels 
as well as a set of orthonormal eigenfunctions are found as
solutions of the Dirac equation. With these orthonormal
eigenfunctions we construct the fermion field operators
and the corresponding Wigner-function operator.
The Wigner function is obtained by taking the ensemble average of
the Wigner-function operator in global thermodynamical equilibrium,
i.e., at constant temperature $T$ and non-zero fermion-number and
chiral-charge chemical potentials $\mu$ and $\mu_5$, respectively. Extracting the vector and axial-vector
components of the Wigner function,
we reproduce the currents of the chiral magnetic and separation effect
in an arbitrarily strong magnetic field. 
\end{abstract}

\preprint{\hfill {\small {ICTS-USTC-17-12}}}

\maketitle

\section{Introduction}

Heavy-ion collisions at ultrarelativistic energies create a new phase of strongly interacting matter, the so-called quark-gluon
plasma (QGP) \cite{Ackermann:2000tr,Adler:2003kt,Adcox:2001jp,Adams:2003kv,Adler:2002tq,Muller:2012zq},
for reviews, see, e.g., 
Refs.~\cite{Rischke:2003mt,Gyulassy:2004zy,Jacobs:2004qv,Jacak:2012dx,Akiba:2015jwa,Rafelski:2015cxa,Koch:2017pda}.
In the QGP, quarks and gluons are deconfined and the chiral symmetry
of the fundamental theory of the strong interaction, quantum chromodynamics
(QCD), is restored. The QGP occurs at temperatures above a deconfinement
and chiral symmetry restoring crossover transition at $T_{\chi}\sim 150$
MeV (for reviews of lattice-QCD calculations, see, e.g., Refs.~\cite{Borsanyi:2010cj,Soltz:2015ula,Ding:2015ona}). 

At temperatures above, but not asymptotically far above, $T_{\chi}$,
the QGP is not a gas of weakly interacting quarks and gluons, but
rather a strongly interacting system, with a surprisingly small
shear viscosity-to-entropy density ratio $\eta/s$ (approaching the value 
estimated from the uncertainty principle \cite{Danielewicz:1984ww}). This leads
to a strong degree of collectivity of the hot and dense system created
in heavy-ion collisions. In fact, the strong collective flow of strongly
interacting matter, parametrized in terms of the elliptic flow coefficient
$v_{2}$ \cite{Ackermann:2000tr,Adler:2003kt,Kolb:2000fha,Teaney:2000cw},
has become the trademark signature of this system: it has been coined
the \textquotedblleft{}most perfect liquid\textquotedblright{} ever
created. 

If transport coefficients, like $\eta/s$, are sufficiently small, 
the system is
close to local thermodynamical equilibrium and 
a fluid-dynamical description for the dynamical evolution of the system
becomes applicable \cite{Kolb:2000fha,Teaney:2000cw,Schenke:2011bn}.
Comparing fluid-dynamical calculations of the collective flow to experimental
data, one has attempted to deduce bounds for the $\eta/s$ ratio \cite{Song:2010mg}.
Such studies indicate that $\eta/s$ could be as small as 0.2, which
is not far from the KSS bound $1/(4\pi)$ or the quantum limit 
suggested by the AdS/CFT correspondence \cite{Kovtun:2004de}. 

However, the dynamics of a heavy-ion collision is complex and influenced
by many effects. The Frankfurt school led by Walter Greiner 
were pioneers in the study of the physics of strong fields in heavy-ion collisions 
\cite{Greiner:1985ce}. It has been recently realized that
the magnetic field created by the moving charges in relativistic heavy-ion
collisions can be of the order of $eB\sim m_{\pi}^{2}$ 
\cite{Kharzeev:2007jp,Skokov:2009qp,Voronyuk:2011jd,Deng:2012pc,Bloczynski:2012en,
McLerran:2013hla,Gursoy:2014aka,Roy:2015coa,Tuchin:2014iua,Li:2016tel}.
While such fields rapidly decay in the vacuum, they can be sustained
for a longer time by an induced current in a conducting medium 
\cite{McLerran:2013hla,Gursoy:2014aka,Tuchin:2014iua,Li:2016tel}.
Such a magnetic field can then lead to an increase of the elliptic
flow \cite{Das:2017qfi,Roy:2017yvg}. In turn, this will increase the value of
$\eta/s$ necessary to describe elliptic-flow data. 

Due to the fact that QCD has a quantum anomaly which gives rise to
an explicit breaking of the $U(1)_{A}$ symmetry, strongly interacting
matter in a magnetic field can also exhibit other interesting effects.
For instance, in case of a local imbalance between right- and left-handed
quarks, a magnetic field leads to a current which separates electric
charges along the direction of the magnetic field, the so-called Chiral
Magnetic Effect (CME) \cite{Vilenkin:1980fu,Kharzeev:2007jp,Fukushima:2008xe,Son:2009tf,Son:2012bg,Son:2012wh,Stephanov:2012ki,Gao:2012ix},
for reviews, see, e.g., Refs.~\cite{Kharzeev:2012ph,Kharzeev:2015znc,Huang:2015oca}.
The CME is associated with the chiral vortical effect (CVE), where an
electric current is induced by the vorticity in a system of charged
particles \cite{Vilenkin:1978hb,Erdmenger:2008rm,Banerjee:2008th,Son:2009tf,Landsteiner:2011cp,Gao:2012ix}.
In anomalous hydrodynamics the CME and CVE must coexist in order to
guarantee the second law of thermodynamics \cite{Son:2009tf,Pu:2010as,Ozonder:2010zy}.
In baryon-rich matter, an axial current is generated which separates
right- and left-handed quarks along the direction of the magnetic
field, the so-called Chiral Separation Effect (CSE) \cite{Kharzeev:2010gr,Landsteiner:2011cp,Gao:2012ix}.
The interplay between CME and CSE leads to so-called Chiral Magnetic
Waves (CMW) \cite{Kharzeev:2010gd,Burnier:2011bf}. The CME has recently
been confirmed in materials such as Dirac and Weyl semi-metals \cite{Son:2012bg,Basar:2013iaa,Li:2014bha}. 

The charged-particle correlations observed in STAR \cite{Abelev:2009ac,Abelev:2009ad}
and ALICE \cite{Abelev:2012pa} experiments are consistent with the
CME prediction. But there were debates that the observed correlations
might arise from other effects such as clustered-particle correlations
\cite{Wang:2009kd} or local charge conservation \cite{Schlichting:2010qia},
so a substantial part of the charged-particle correlation measured
in experiments may come from background effects. Recently, the CMS collaboration
has measured the charged-particle correlations in pPb collisions
\cite{Khachatryan:2016got} and found a result similar to that of
STAR \cite{Abelev:2009ac,Abelev:2009ad} and ALICE \cite{Abelev:2012pa}
in AuAu and PbPb collisions. The CMS result indicates that the measured
azimuthal correlation of charged particles at TeV energies may
be a background effect. Further theoretical and experimental
investigations are needed to separate the signal from the background 
\cite{Sorensen:2017,Hirono:2014oda,Deng:2016knn}. 

The covariant Wigner-function method 
\cite{Heinz:1983nx,Elze:1986hq,Elze:1986qd,Vasak:1987um,Elze:1989un,BialynickiBirula:1993um,Abada:1996hq} 
for spin-1/2 fermions is a useful tool to study the CME, CVE, and other
related effects \cite{Gao:2012ix,Chen:2012ca,Son:2012wh,Gao:2015zka,Fang:2016vpj,Hidaka:2016yjf,Mueller:2017arw}.
However, previous investigations of these phenomena rely on the assumption that the
magnetic field is weak and can be treated as a perturbation. The purpose
of the present work is to show that the Wigner-function method can also be applied
for magnetic fields of arbitrary strength. To this
end, we will derive the exact solution for the fermion Wigner function
in a constant, arbitrarily strong magnetic field $\mathbf{B}$
in an extended system in global thermodynamical equilibrium, i.e., at constant 
temperature $T$ and fermion-number chemical potential $\mu$. In
order to study the CME and CSE, we also allow for a non-zero chiral-charge
chemical potential $\mu_{5}$, i.e., we can independently control
the number densities of right- and left-handed fermions through their
associated chemical potentials $\mu_{L,R}\sim\mu\mp\mu_{5}$. We will
confirm that the CME and the CSE are natural consequences in such a system. 

We also note that the covariant Wigner-function method has also been applied to 
derive the kinetic equation for gluons in the background fields by one 
of the authors in collaboration with Walter Greiner \cite{Wang:2001dm,Wang:2002qe}.  

This paper is organized as follows. For determining the Wigner function
we need to compute grand canonical ensemble averages of two fermion field 
operators. The ensemble averages require, in turn, a complete
set of basis functions with which one can compute the Gibbs operator-weighted
traces. Since these traces become simple if one diagonalizes the Hamilton
operator (including fermion-number and chiral-charge chemical potential)
of the system, we first derive in Sec.~\ref{sec:Landau-Levels} the
exact solution of the one-particle Dirac equation in the presence
of a constant magnetic field and chemical potentials $\mu,\mu_{5}$,
i.e., we find the corresponding energy eigenvalues and wave functions.
It turns out that this can be done in completely analytical form.
We will see that the original expression for the energy of the Landau
levels is modified in the presence of non-zero $\mu_{5}$. Then, in
Sec.~\ref{sec:Field-operators,-Hamilton}, we construct the fermion
field operators as an expansion in terms of the exact solutions of
the Dirac equation derived in the previous section. The Hamilton operator
is diagonal in this basis, i.e., it only contains single-particle
creation and annihilation operators. We then compute the Wigner function
in Sec.~\ref{sec:Wigner-Functions}. The latter possesses a decomposition
in terms of the independent generators of the Clifford algebra, the
so-called Dirac-Heisenberg-Wigner functions. From these, we then derive
expressions for the fermion-number and chiral-charge currents in Sec.~\ref{sec:Chiral-Effects}
and recover the results derived previously in the weak-field limit,
which give rise to the CME and CSE.

We take fermions to have positive charge $Q=+e$
and the magnetic field to point in the $z$-direction. We use the following notations for four-vectors:
$X=(x^{\mu})=(t,\mathbf{r})=(t,x,y,z)$, $P=(p^{\mu})=(E,\mathbf{p})=(E,p_{x},p_{y},p_{z})$.
We choose the temporal gauge $A_{0}=0$ throughout this paper.

\section{Landau levels and wave functions}

\label{sec:Landau-Levels}The momentum spectrum of a free particle
is continuous in an infinite volume. In a constant magnetic field,
the longitudinal momentum along the field remains continuous while
the transverse momentum becomes discrete. The dispersion relation is 
\begin{eqnarray}
E_{p_{z}}^{(n)} & = & \sqrt{m^{2}+p_{z}^{2}+2neB}.\label{eq:LL-en}
\end{eqnarray}
Here $n=0,1,2,\cdots$ labels the Landau energy levels \cite{Landau:1977}
[see also the recent review \cite{Miransky:2015ava}], characterizing
the quantization of transverse momentum. In Eq.~(\ref{eq:LL-en}) 
the quantum number $n$ can also be written as
$n=n^{\prime}+ 1/2 +s$, with $s=\pm 1/2$ being the spin of the fermion 
and $n^{\prime}=0,1,2,\cdots $ being the orbital quantum number. 
The number of states for fixed $p_{z}$ is $|eB|L_{x}L_{y}/(2\pi)$ at $n=0$ and
$|eB|L_{x}L_{y}/\pi$ at $n>0$, where $L_{x}L_{y}$ is the
transverse area of the system. In this section we will solve the Dirac
equation including space-time independent chemical potentials for
both fermion number and chiral charge and show how the dispersion
relation (\ref{eq:LL-en}) is modified. We will also derive the corresponding wave functions.

\subsection{Dirac equation for massive particles and Landau levels\label{sub:Dirac-equation-for}}

We use the Weyl (or chiral) representation for the Dirac matrices,
\begin{eqnarray}
\gamma^{\mu} & = & \left(\begin{array}{cc}
0 & \sigma^{\mu}\\
\bar{\sigma}^{\mu} & 0
\end{array}\right),
\end{eqnarray}
with $\sigma^{\mu}=(\boldsymbol{1},\boldsymbol{\sigma})$ and 
$\bar{\sigma}^{\mu}=(\boldsymbol{1},-\boldsymbol{\sigma})$.
In this representation, $\gamma^{5}=i\gamma^{0}\gamma^{1}\gamma^{2}\gamma^{3}
=\mathrm{diag}(-\boldsymbol{1},\boldsymbol{1})$.
Including the fermion-number and chiral-charge chemical potentials $\mu$
and $\mu_{5}$ or equivalently the chemical potentials for left- and
right-handed chirality $\mu_{L,R}$, the Dirac Lagrangian in the presence of an
external electromagnetic field $A^{\sigma}$ is 
\begin{equation}
\mathcal{L}=\bar{\psi}[i\gamma^{\sigma}(\partial_{\sigma}+ieA_{\sigma})-m+\mu\gamma^{0}
+\mu_{5}\gamma^{0}\gamma_{5}]\psi.
\end{equation}
The corresponding Hamilton density is %
\begin{eqnarray}
\mathcal{H} & = & i\psi^{\dagger}\partial_{t}\psi-\mathcal{L}\nonumber \\
 & = & \psi^{\dagger}\left[\boldsymbol{\alpha}\cdot(-i\boldsymbol{\nabla}-e\mathbf{A})+m\gamma^{0}
 -\mu-\mu_{5}\gamma_{5}\right]\psi,\label{eq:ham}
\end{eqnarray}
where $\boldsymbol{\alpha}=\gamma^{0}\boldsymbol{\gamma}
=\mathrm{diag}(-\boldsymbol{\sigma},\boldsymbol{\sigma})$.
The Dirac equation reads 
\begin{eqnarray}
[i\gamma^{\sigma}(\partial_{\sigma}+ieA_{\sigma})-m+\mu\gamma^{0}+\mu_{5}\gamma^{0}\gamma_{5}]\psi(x) 
& = & 0.\label{eq:dirac-eq}
\end{eqnarray}
It can be rewritten in the form of a Schr\"odinger equation, 
\begin{equation}
i\frac{\partial\psi}{\partial t}=[\boldsymbol{\alpha}\cdot(-i\boldsymbol{\nabla}-e\mathbf{A})
+m\gamma^{0}-\mu-\mu_{5}\gamma_{5}]\psi,\label{eq:dirac-eq-2}
\end{equation}
where we can read off the Hamilton operator for a Dirac particle,
\begin{equation}
\hat{H}=\boldsymbol{\alpha}\cdot(-i\boldsymbol{\nabla}-e\mathbf{A})+m\gamma^{0}-\mu-\mu_{5}\gamma_{5}.
\end{equation}
In the Landau gauge, a constant and homogeneous external magnetic field  
pointing in the $z$-direction and the associated vector potential are given by 
\begin{eqnarray}
\mathbf{B} & = & B\mathbf{e}_{z},\nonumber \\
\mathbf{A} & = & -By\mathbf{e}_{x}.\label{eq:vector-pot}
\end{eqnarray}
Without loss of generality, we take $B>0$. Of course one can also
choose a symmetric form for the vector potential, $\mathbf{A}=\frac{1}{2}\mathbf{B}\times\mathbf{r}$.
The Wigner function derived in Sec.~\ref{sec:Wigner-Functions} will not depend on the choice of gauge.

Since $\hat{H}$ does not depend on $x$ and $z$, $\frac{\partial}{\partial x}$ and $\frac{\partial}{\partial z}$
commute with the Hamilton operator, $[\frac{\partial}{\partial x},\hat{H}]=[\frac{\partial}{\partial z},\hat{H}]=0$. 
This indicates
that $p_{x}$ and $p_{z}$ are conserved quantities. 
Thus, the solution to Eq.~(\ref{eq:dirac-eq-2}) can be cast into the following form 
\begin{eqnarray}
\psi(t,\mathbf{r}) & = & e^{-iEt+ip_{x}x+ip_{z}z}\xi(p_{x},p_{z},y).\label{eq:solution}
\end{eqnarray}

We can make the decomposition 
\begin{equation}
\xi(p_{x},p_{z},y)=\left(\begin{array}{c}
\chi_{L}(p_{x},p_{z},y)\\
\chi_{R}(p_{x},p_{z},y)
\end{array}\right),
\end{equation}
where $\chi_{L,R}$ are Pauli spinors for left- and right-handed chirality,
respectively. Then Eq.~(\ref{eq:dirac-eq-2}) can be simplified as
\begin{eqnarray}
\left[E+\mu-\mu_{5}+\sigma_{3}p_{z}+\sigma_{1}(p_{x}+eBy)-i\sigma_{2}\frac{\partial}{\partial y}\right]\chi_{L} 
& = & m\chi_{R},\nonumber \\
\left[E+\mu+\mu_{5}-\sigma_{3}p_{z}-\sigma_{1}(p_{x}+eBy)+i\sigma_{2}\frac{\partial}{\partial y}\right]\chi_{R} 
& = & m\chi_{L}.
\end{eqnarray}
Using the standard form of the Pauli matrices this becomes 
\begin{eqnarray}
\left(\begin{array}{cc}
E+\mu-\mu_{5}+p_{z} & \sqrt{2eB}\hat{a}^{\dagger}\\
\sqrt{2eB}\hat{a} & E+\mu-\mu_{5}-p_{z}
\end{array}\right)\chi_{L} & = & m\chi_{R},\nonumber \\
\left(\begin{array}{cc}
E+\mu+\mu_{5}-p_{z} & -\sqrt{2eB}\hat{a}^{\dagger}\\
-\sqrt{2eB}\hat{a} & E+\mu+\mu_{5}+p_{z}
\end{array}\right)\chi_{R} & = & m\chi_{L},\label{eq:eq-R,L}
\end{eqnarray}
where we have introduced the operators,
\begin{eqnarray}
\hat{a} & = & \frac{1}{\sqrt{2eB}}\left[\frac{\partial}{\partial y}+eB\left(y+\frac{p_{x}}{eB}\right)\right],\nonumber \\
\hat{a}^{\dagger} & = & \frac{1}{\sqrt{2eB}}\left[-\frac{\partial}{\partial y}+eB\left(y+\frac{p_{x}}{eB}\right)\right],
\end{eqnarray}
which are the annihilation and creation operators for a harmonic oscillator with mass $m$ and 
frequency $m\omega=\sqrt{eB}$ and centered at $-p_{x}/(eB)$.
It is straightforward to check that $[\hat{a},\hat{a}^{\dagger}]=1$.
Eliminating $\chi_{L}$ or $\chi_{R}$ from Eq.~(\ref{eq:eq-R,L})
we can derive equations for $\chi_{R}$ and $\chi_{L}$, 
\begin{eqnarray}
\left(\begin{array}{cc}
(E+\mu)^{2}-\Lambda^{-} & 2\mu_{5}\sqrt{2eB}\hat{a}^{\dagger}\\
2\mu_{5}\sqrt{2eB}\hat{a} & (E+\mu)^{2}-\Lambda^{+}
\end{array}\right)\chi_{R,L}(p_{x},p_{z},y) & = & 0,\label{eq:eq-chi-R}
\end{eqnarray}
where we defined the operators 
\begin{equation}
\Lambda^{\pm}=m^{2}+(p_{z}\pm\mu_{5})^{2}+2eB\left(\hat{a}^{\dagger}\hat{a}+\frac{1}{2}\right)\pm eB.
\end{equation}

In order to solve Eq.~(\ref{eq:eq-chi-R}), we expand $\chi_{R,L}(p_{x},p_{z},y)$
in a basis of eigenfunctions of the harmonic oscillator, $\phi_{n}(p_{x},y)$
\cite{Landau:1977}, 
\begin{eqnarray}
\chi_{R,L}(p_{x},p_{z},y) & = & \sum_{n=0}^{\infty}\left(\begin{array}{c}
c_{n}(p_{x},p_{z})\\
d_{n}(p_{x},p_{z})
\end{array}\right)\phi_{n}(p_{x},y),\label{eq:chi-R}
\end{eqnarray}
where $c_{n}$ and $d_{n}$ depend on $p_{x}$ and $p_{z}$,
and $\phi_{n}(p_{x},y)$ are given by 
\begin{eqnarray}
\phi_{n}(p_{x},y) & = & \left(\frac{eB}{\pi}\right)^{1/4}\frac{1}{\sqrt{2^{n}n!}}\exp\left[-\frac{eB}{2}
\left(y+\frac{p_{x}}{eB}\right)^{2}\right]H_{n}\left[\sqrt{eB}\left(y+\frac{p_{x}}{eB}\right)\right],\label{eq:wave}
\end{eqnarray}
where $H_{n}$ is the $n$-th Hermite polynomial. 
Note that the eigenfunctions $\phi_{n}(p_{x},y)$ do not depend on $p_x$ and $y$ separately,
but only on the linear combination $y + p_x/(eB)$, i.e., $\phi_n(p_x,y) \equiv \phi_n(y-y_0)$,
where $y_0 \equiv - p_x/(eB)$. The interpretation is that, for given
$p_x$, the $y$ coordinate of the center of the Landau orbit is precisely determined by $y_0$.
One can show \cite{Landau:1977} that also its $x$ coordinate, $x_0 = x+p_y/(eB)$, 
can be precisely determined, since $\hat{x}_0 = x -i/(eB) \partial/\partial y$ 
commutes with $\hat{H}$. However, a simultaneous
determination of $y_0$ and $x_0$ is not possible, 
since $\hat{x}_0$ and $\hat{y}_0 = - i/(eB) \partial/\partial x$ 
do not commute with each other. 
While the probability to find a particle at a given
$y$ coordinate is determined by the wave function $\phi_{n}(p_{x},y)$, 
due to Heisenberg's uncertainty principle 
its $x$ coordinate remains completely undetermined, since $p_x$ is a good quantum number.
Despite the fact that $\phi_n$ depends on $y-y_0$ only, 
in the following we keep the notation $\phi_{n}(p_{x},y)$, because
when we discuss second quantization in Sec.~\ref{sec:Field-operators,-Hamilton}, we need a label ($p_x$) to 
keep track of the Landau orbit in which a particle is created or annihilated.

The eigenfunctions $\phi_{n}(p_{x},y)$
satisfy the orthonormality condition 
\begin{equation} \label{eq:orthophi}
\int dy\phi_{n}(p_{x},y)\phi_{n^{\prime}}(p_{x},y)=\delta_{nn^{\prime}}.
\end{equation}
Furthermore, applying annihilation and creation operators, 
\begin{eqnarray}
\hat{a}\phi_{n}(p_{x},y)& =& \sqrt{n}\; \phi_{n-1}(p_{x},y), \nonumber \\
\hat{a}^{\dagger}\phi_{n}(p_{x},y)& =& \sqrt{n+1}\; \phi_{n+1}(p_{x},y), 
\end{eqnarray}
where $n\geq 0$ and we assumed $\phi_{-1}=0$. Inserting the above expansion
into Eq.~(\ref{eq:eq-chi-R}), we obtain 
\begin{eqnarray}
\sum_{n=0}^{\infty}\left(\begin{array}{cc}
[(E+\mu)^{2}-\lambda_{n}^{-}]\phi_{n} & 2\mu_{5}\sqrt{2(n+1)eB}\phi_{n+1}\\
2\mu_{5}\sqrt{2neB}\phi_{n-1} & [(E+\mu)^{2}-\lambda_{n+1}^{+}]\phi_{n}
\end{array}\right)\left(\begin{array}{c}
c_{n}\\
d_{n}
\end{array}\right) & = & 0,
\end{eqnarray}
where
\begin{eqnarray}
\lambda_{n}^{\pm} & = & m^{2}+(p_{z}\pm\mu_{5})^{2}+2neB.
\end{eqnarray}

Using the orthonormality condition (\ref{eq:orthophi}) we can derive
equations for $c_{n}$ and $d_{n}$, 
\begin{eqnarray}
[(E+\mu)^{2}-\lambda_{0}^{-}]c_{0} & = & 0,\nonumber \\
{}[(E+\mu)^{2}-\lambda_{n}^{-}]c_{n} & = & -2\mu_{5}\sqrt{2neB}d_{n-1},\ \ \ n>0,\nonumber \\
{}[(E+\mu)^{2}-\lambda_{n}^{+}]d_{n-1} & = & -2\mu_{5}\sqrt{2neB}c_{n},\ \ \ n>0.\label{eq:cndn}
\end{eqnarray}
We observe that $c_{0}$ decouples from the other coefficients while
the $c_{n}$ $(n>0)$ always couple to $d_{n-1}$ $(n>0)$. The energy
eigenvalue for $n=0$ is obtained by demanding a non-zero value for
$c_{0}$. Then the first equation (\ref{eq:cndn}) gives $E=\pm E_{p_{z}}^{(0)}-\mu$
for positive/negative-energy states, where 
\begin{equation}
E_{p_{z}}^{(0)}=\sqrt{m^{2}+(p_{z}-\mu_{5})^{2}}\label{eq:en-lowest}
\end{equation}
is the energy of the lowest Landau level. A non-zero $c_{0}$ means
that this level is occupied by a fermion with spin up. 

The energy eigenvalues $E$ for $n>0$ are obtained by decoupling
the second and third equations in Eq.~(\ref{eq:cndn}). For positive/negative-energy
states these eigenvalues are given by $E=\pm E_{p_{z}s}^{(n)}-\mu$,
where 
\begin{eqnarray}
E_{p_{z}s}^{(n)} & {\normalcolor =} & \sqrt{m^{2}+\left(\sqrt{p_{z}^{2}+2neB}-s\mu_{5}\right)^{2}}\label{eq:eigen-energy}
\end{eqnarray}
is the energy of a Landau level for $n>0$ and $s=\pm1$ (helicity
in the massless case) \cite{Fukushima:2008xe}. We note that the energy
levels depend on $p_{z}$, $n$, and $s$ and are independent of $p_{x}$.
One also observes that the two-fold degeneracy of the conventional
Landau levels (\ref{eq:LL-en}) with respect to spin (or helicity)
is now lifted by a non-zero $\mu_{5}$. 

In the case of vanishing chiral-charge chemical potential, $\mu_{5}=0$,
all coefficients $c_{n}$, $d_{n}$ decouple from each other. Equation
(\ref{eq:cndn}) becomes 
\begin{eqnarray}
[(E+\mu)^{2}-(m^{2}+p_{z}^{2}+2neB)]c_{n} & = & 0,\nonumber \\
\left\{ (E+\mu)^{2}-[m^{2}+p_{z}^{2}+2(n+1)eB]\right\} d_{n} & = & 0,\label{eq:eq-dn}
\end{eqnarray}
for $n\geq0$, and we obtain the conventional Landau energy levels
(\ref{eq:LL-en}). The lowest Landau level with $n=0$ is occupied
by a fermion with spin up. The higher Landau levels are two-fold degenerate,
being occupied by fermions with spin up and spin down. 

We conclude this subsection with some remarks on the degeneracy of the energy levels 
in the Landau gauge (\ref{eq:vector-pot}). 
In this gauge a quantum state is labeled by a set of quantum numbers $\{n,s,p_x,p_z\}$, so 
the sum over quantum states for a function $\mathcal{F}$ is $\sim 
L_xL_z \sum_{n,s} \int \frac{dp_{x}}{2\pi}\frac{dp_{z}}{2\pi} \mathcal{F}(n,s,p_x,p_z)$.  
If $\mathcal{F}$ does not depend on $p_x$, we can trivially perform the integral over $p_x$.
The integral is bounded by the requirement that
the Landau orbit labelled by $p_x$ is still located inside the transverse area $L_x L_y$, i.e.,
by the requirement $0 \leq y_0 \leq L_y$ (where we neglected the small radius of the orbit with respect 
to $L_y$). Thus, $0 \leq p_x/(eB)\leq L_y$, and the degeneracy factor becomes 
$L_x\int \frac{p_{x}}{2\pi}= \frac{e\Phi}{2\pi}$ with $\Phi \equiv BL_xL_y$ being the magnetic flux 
through the transverse area $L_x L_y$ \cite{Landau:1977}.

\subsection{Landau wave functions}

We now determine the eigenspinors $\chi_{s}^{(n)}(p_{x},p_{z},y)$
associated with the Landau energy levels (\ref{eq:eigen-energy}).
We start by assuming that the energy is equal to that of the lowest
Landau level $E_{p_{z}}^{(0)}$. From Eq.~(\ref{eq:cndn}) we conclude
that $c_{0}$ can be non-zero while all other coefficients $c_{n},d_{n-1}$
with $n>0$ have to vanish. The normalized eigenspinor associated
with this state is 
\begin{equation}
\chi^{(0)}(p_{x},y)=\left(\begin{array}{c}
1\\
0
\end{array}\right)\phi_{0}(p_{x},y).\label{eq:chi-0}
\end{equation}
Now assume that the energy is equal to $E_{p_{z}s}^{(n)}$ with $n>0$,
cf.\ Eq.~(\ref{eq:eigen-energy}). Then only the coefficients $c_{n}$,
$d_{n-1}$ can be non-zero while all other coefficients $c_{m}$,
$d_{m-1}$ with $m\neq n$ have to vanish. The normalized eigenspinor
associated with this Landau level is 
\begin{eqnarray}
\chi_{s}^{(n)}(p_{x},p_{z},y) & = & \frac{1}{\sqrt{2\sqrt{p_{z}^{2}+2neB}}}\left(\begin{array}{c}
\sqrt{\sqrt{p_{z}^{2}+2neB}+sp_{z}}\phi_{n}(p_{x},y)\\
s\sqrt{\sqrt{p_{z}^{2}+2neB}-sp_{z}}\phi_{n-1}(p_{x},y)
\end{array}\right).\label{eq:chi-1}
\end{eqnarray}
This result is obtained as follows. Since for given energy $E_{p_{z}s}^{(n)}$
only the coefficients $c_{n}$ and $d_{n-1}$ are non-zero, we conclude
from Eq.~(\ref{eq:eq-chi-R}) that 
\begin{equation}
\chi_{s}^{(n)}(p_{x},p_{z},y)=\left(\begin{array}{c}
c_{n}(p_{x},p_{z})\phi_{n}(p_{x},y)\\
d_{n-1}(p_{x},p_{z})\phi_{n-1}(p_{x},y)
\end{array}\right).\label{eq:cn-dn-1}
\end{equation}
In order to fulfill the second Eq.~(\ref{eq:cndn}) we have to demand
that $c_{n}\sim-2\mu_{5}\sqrt{2neB}$ while $d_{n-1}\sim(E_{p_{z}s}^{(n)})^{2}-\lambda_{n}^{-}$.
Inserting this into Eq.~(\ref{eq:cn-dn-1}) and normalizing the eigenspinor
yields Eq.~(\ref{eq:chi-1}). For later use, we then define $c_{n}(p_{x},p_{z})$
and $d_{n-1}(p_{x},p_{z})$ by the values in Eq.~(\ref{eq:chi-1}). 

The eigenspinors (\ref{eq:chi-0}) and (\ref{eq:chi-1}) fulfill the
following orthonormality conditions 
\begin{eqnarray}
\int dy\chi^{(0)\dagger}(p_{x},y)\chi^{(0)}(p_{x},y) & = & 1,\nonumber \\
\int dy\chi^{(0)\dagger}(p_{x},y)\chi_{s}^{(n)}(p_{x},p_{z},y) & = & 0,\nonumber \\
\int dy\chi_{s}^{(n)\dagger}(p_{x},p_{z},y)\chi_{s^{\prime}}^{(n^{\prime})}(p_{x},p_{z},y) 
& = & \delta_{nn^{\prime}}\delta_{ss^{\prime}}.
\end{eqnarray}
From Eq.~(\ref{eq:solution}) we obtain the Dirac wave functions corresponding
to the various Landau levels, 
\begin{eqnarray}
\psi_{r}^{(0)}(t,\mathbf{r}) & = & \exp[-irE_{p_{z}}^{(0)}t+i\mu t+ip_{x}x+ip_{z}z]\xi_{r}^{(0)}(p_{x},p_{z},y),\nonumber \\
\psi_{rs}^{(n)}(t,\mathbf{r}) & = & \exp[-irE_{p_{z}s}^{(n)}t+i\mu t+ip_{x}x+ip_{z}z]\xi_{rs}^{(n)}(p_{x},p_{z},y),
\label{eq:eigen-function}
\end{eqnarray}
where $r=\pm$ denotes positive- or negative-energy states and $s$
denotes the helicity of the state. Here the Dirac spinors $\xi_{r}^{(0)}$
and $\xi_{rs}^{(n)}$, which depend on momentum $p_{x}$, $p_{z}$,
and $y$, are defined by 
\begin{eqnarray}
\xi_{r}^{(0)}(p_{x},p_{z},y) & = & \frac{1}{\sqrt{2E_{p_{z}}^{(0)}}}\left(\begin{array}{c}
r\sqrt{E_{p_{z}}^{(0)}-r(p_{z}-\mu_{5})}\\
\sqrt{E_{p_{z}}^{(0)}+r(p_{z}-\mu_{5})}
\end{array}\right)\otimes\chi^{(0)}(p_{x},y),\nonumber \\
\xi_{rs}^{(n)}(p_{x},p_{z},y) & = & \frac{1}{\sqrt{2E_{p_{z}s}^{(n)}}}\left(\begin{array}{c}
r\sqrt{E_{p_{z}s}^{(n)}+r\mu_{5}-rs\sqrt{p_{z}^{2}+2neB}}\\
\sqrt{E_{p_{z}s}^{(n)}-r\mu_{5}+rs\sqrt{p_{z}^{2}+2neB}}
\end{array}\right)\otimes\chi_{s}^{(n)}(p_{x},p_{z},y). 
\label{eq:xi-rs}
\end{eqnarray}
We can easily check that the quantities under the roots in 
Eq.~(\ref{eq:xi-rs}) have non-negative values because 
$E_{p_{z}}^{(0)}\geq\mid p_{z}-\mu_{5}\mid$ 
and $E_{p_{z}s}^{(n)}\geq\left|\sqrt{p_{z}^{2}+2neB}-s\mu_{5}\right|$. 
The first equation is obtained by using Eq.~(\ref{eq:eq-R,L}) to
express $\chi_{L,R}$ in terms of $\chi_{R,L}$, assuming $\chi_{R,L}\sim\chi^{(0)}$,
and normalizing the resulting Dirac spinor. The second equation is
obtained analogously, assuming $\chi_{R,L}\sim\chi_{s}^{(n)}$. The
Dirac spinors (\ref{eq:xi-rs}) satisfy the following orthonormality relations, 
\begin{eqnarray}
\int dy\xi_{r}^{(0)\dagger}(p_{x},p_{z},y)\xi_{r^{\prime}}^{(0)}(p_{x},p_{z},y) & = & \delta_{rr^{\prime}},\nonumber \\
\int dy\xi_{r}^{(0)\dagger}(p_{x},p_{z},y)\xi_{r^{\prime}s}^{(n)}(p_{x},p_{z},y) & = & 0,\nonumber \\
\int dy\xi_{rs}^{(n)\dagger}(p_{x},p_{z},y)\xi_{r^{\prime}s^{\prime}}^{(n^{\prime})}(p_{x},p_{z},y) 
& = & \delta_{nn^{\prime}}\delta_{rr^{\prime}}\delta_{ss^{\prime}}.\label{eq:orthogonality}
\end{eqnarray}

\section{Field operators, Hamilton operator, and distribution function\label{sec:Field-operators,-Hamilton}}

\subsection{Field operators}

As shown in the last section, the Dirac eigenspinors (\ref{eq:eigen-function})
form an orthonormal basis and can thus be used in an expansion of the
fermion field operator, 
\begin{eqnarray}
\psi(t,\mathbf{r}) & = & e^{i\mu t}\sum_{n,s}\int_{p_{x},p_{z}}\left\{ a_{p_{x}p_{z}s}^{(n)}\; \xi_{+,s}^{(n)}(p_{x},p_{z},y)
\exp\left[-iE_{p_{z}s}^{(n)}t+ip_{x}x+ip_{z}z\right]\right.\nonumber \\
 &  & \hspace*{2cm} \left.+b_{-p_{x},-p_{z},s}^{(n)\dagger}\;\xi_{-,s}^{(n)}(p_{x},p_{z},y)
 \exp\left[iE_{p_{z}s}^{(n)}t+ip_{x}x+ip_{z}z\right]\right\} .\label{eq:field-operator-1}
\end{eqnarray}
Here, $a_{p_{x}p_{z}s}^{(n)}$ is the annihilation operator for fermions
with momentum $p_{x}$, $p_{z}$ in the Landau level $E_{p_{z}s}^{(n)}$
and $b_{-p_{x},-p_{z},s}^{(n)\dagger}$ is the creation operator for
anti-fermions with momentum $-p_{x}$, $-p_{z}$ in the same Landau
level. We also defined $\int_{p_{x},p_{z}}\equiv\int dp_{x}dp_{z}/(2\pi)^{2}$ and 
\begin{eqnarray}
\sum_{n,s}f_{s}^{(n)} & \equiv & f^{(0)}+\sum_{n>0,s=\pm}f_{s}^{(n)}
\end{eqnarray}
for any function $f_{s}^{(n)}$ which depends on the Landau level
$n$ and the helicity $s$. The first term is for the lowest Landau
level with $n=0$, which is always occupied by a fermion/anti-fermion
with spin up/down. We see that the chemical potential $\mu$ contributes
only a global phase factor $e^{i\mu t}$ to the field. 

Note again
that the momentum variable $p_x$ serves as a label for the individual Landau
orbits [with center located at $y_0 = - p_x/(eB)$]. The integral over $p_x$ can
thus also be interpreted as a (continuous) summation over these Landau levels.
However, we cannot trivially perform the $p_x$ integral (giving rise to the well-known
degeneracy factor $\frac{e\Phi}{2\pi}$), because
the integrand in Eq.\ (\ref{eq:field-operator-1}) depends on $p_x$: each Landau orbit
has its own Fock space on which the annihilation and creation 
$a_{p_{x}p_{z}s}^{(n)}, b_{-p_{x},-p_{z},s}^{(n)\dagger}$ act, and each wavefunction of the 
respective particle needs to be associated to the particular Landau orbit (labelled by $p_x$) where
it was annihilated or created.

We assume that all operators satisfy the following anti-commutation
relations 
\begin{eqnarray}
\left\{ a_{p_{x}p_{z}s}^{(n)},a_{q_{x}q_{z}s^{\prime}}^{(n^{\prime})\dagger}\right\}  
& = & (2\pi)^{2}\delta(p_{x}-q_{x})\delta(p_{z}-q_{z})\delta_{nn^{\prime}}\delta_{ss^{\prime}},\nonumber \\
\left\{ b_{p_{x}p_{z}s}^{(n)},b_{q_{x}q_{z}s^{\prime}}^{(n^{\prime})\dagger}\right\}  
& = & (2\pi)^{2}\delta(p_{x}-q_{x})\delta(p_{z}-q_{z})\delta_{nn^{\prime}}\delta_{ss^{\prime}},\label{eq:op-comm}
\end{eqnarray}
while all other anti-commutators vanish. Then one can verify the following equal-time
anticommutation relations for the field operators, 
\begin{eqnarray}
\left\{ \psi_{\alpha}(t,\mathbf{r}),\psi_{\beta}^{\dagger}(t,\mathbf{r}^{\prime})\right\}  
& = & \delta_{\alpha\beta}\delta^{(3)}(\mathbf{r}-\mathbf{r}^{\prime}),\nonumber \\
\left\{ \psi_{\alpha}(t,\mathbf{r}),\psi_{\beta}(t,\mathbf{r}^{\prime})\right\}  
& = & \left\{ \psi_{\alpha}^{\dagger}(t,\mathbf{r}),\psi_{\beta}^{\dagger}(t,\mathbf{r}^{\prime})\right\} =0.\label{eq:field-comm}
\end{eqnarray}

\subsection{Hamilton operator for Dirac fields and distribution functions }

Integrating the Hamilton density (\ref{eq:ham}) over space, using
Eq.~(\ref{eq:field-operator-1}) as well as the orthonormality relations
(\ref{eq:orthogonality}) and anticommutation relations (\ref{eq:op-comm}), (\ref{eq:field-comm}),
we obtain the Hamilton operator in the presence of fermion-number
and chiral-charge chemical potentials, 
\begin{eqnarray}
\hat{H} & = & \sum_{n,s}\int_{p_{x},p_{z}}\left[(E_{p_{z}s}^{(n)}-\mu)\hat{n}_{p_{x}p_{z}s}^{(n)}
+(E_{-p_{z}s}^{(n)}+\mu)\left(\hat{\bar{n}}_{p_{x}p_{z}s}^{(n)}-1\right)\right].\label{eq:ham-number}
\end{eqnarray}
In the above equation, $\hat{n}_{p_{x}p_{z}s}^{(n)}=a_{p_{x}p_{z}s}^{(n)\dagger}a_{p_{x}p_{z}s}^{(n)}$
and $\hat{\bar{n}}_{p_{x}p_{z}s}^{(n)}=b_{p_{x}p_{z}s}^{(n)\dagger}b_{p_{x}p_{z}s}^{(n)}$
are the number operators for particles and anti-particles, respectively. 
We note that the vacuum energy (without matter) in an electromagnetic field has been 
calculated by Heisenberg and Euler in 1935 \cite{Heisenberg:1935qt} 
and by Weisskopf in 1936 \cite{Weisskopf:1996bu}. Since the occupation numbers 
(being the expectation values of the respective occupation number operators) do not depend on $p_x$ explicitly,
cf.\ Eqs.\ (\ref{eq:part-n}) and (\ref{eq:ant-part-n}), the integral over $p_x$ 
can be performed trivially and, as already mentioned above, gives the well-known degeneracy factor $e\Phi/(2\pi)$.

With the Hamilton operator (\ref{eq:ham-number}), the grand partition
function of the system at temperature $T=\beta^{-1}$ reads 
\begin{eqnarray}
\varXi & = & \mathrm{Tr}\left[\exp\left(-\beta\hat{H}\right)\right]\nonumber \\
 & = & \mathrm{Tr}\left\{ \exp\left[-\beta\sum_{n,s}\int_{p_{x},p_{z}}(E_{p_{z}s}^{(n)}-\mu)\hat{n}_{p_{x}p_{z}s}^{(n)}
 -\beta\sum_{n,s}\int_{p_{x},p_{z}}(E_{-p_{z}s}^{(n)}+\mu)\left(\hat{\bar{n}}_{p_{x}p_{z}s}^{(n)}-1\right)\right]\right\} .
\end{eqnarray}
The grand canonical ensemble average of an operator $\hat{O}$ is
given by
\begin{eqnarray}
\left\langle \hat{O}\right\rangle  & = & \frac{1}{\varXi}\mathrm{Tr}\left[\hat{O}\exp\left(-\beta\hat{H}\right)\right].
\end{eqnarray}
In a similar way we can calculate the average particle number in the
state with quantum numbers $(n,s,p_{x},p_{z})$, 
\begin{eqnarray}
\left\langle \hat{n}_{p_{x}p_{z}s}^{(n)}\right\rangle  & = & \frac{\mathrm{Tr}_{R}\left\{ \hat{n}_{p_{x}p_{z}s}^{(n)}
\exp\left[-\beta(E_{p_{z}s}^{(n)}-\mu)\hat{n}_{p_{x}p_{z}s}^{(n)}\right]\right\} }{
\mathrm{Tr}_{R}\left\{ \exp\left[-\beta(E_{p_{z}s}^{(n)}-\mu)\hat{n}_{p_{x}p_{z}s}^{(n)}\right]\right\} },
\end{eqnarray}
where the reduced trace is defined as 
\begin{eqnarray}
\mathrm{Tr}_{R}(\hat{O}) & = & \left\langle 0\right|\hat{O}\left|0\right\rangle 
+\left\langle 0\right|a_{p_{x}p_{z}s}^{(n)}\hat{O}a_{p_{x}p_{z}s}^{(n)\dagger}\left|0\right\rangle .
\end{eqnarray}
Here $\left|0\right\rangle $ is the vacuum state and $a_{p_{x}p_{z}s}^{(n)\dagger}\left|0\right\rangle $
is the one-particle state. States with more than one particle with
the same quantum numbers do not exist due to the Pauli principle.
Then it is straightforward to obtain 
\begin{eqnarray}
\left\langle \hat{n}_{p_{x}p_{z}s}^{(n)}\right\rangle  & = & \frac{1}{\exp\left[\beta(E_{p_{z}s}^{(n)}-\mu)\right]+1}
=f_{\mathrm{FD}}(E_{p_{z}s}^{(n)}-\mu).\label{eq:part-n}
\end{eqnarray}
In the same way we obtain 
\begin{eqnarray}
\left\langle \hat{\bar{n}}_{p_{x}p_{z}s}^{(n)}\right\rangle  & = & f_{\mathrm{FD}}(E_{-p_{z}s}^{(n)}+\mu).\label{eq:ant-part-n}
\end{eqnarray}
We see that the number distributions of particles and anti-particles
follow Fermi-Dirac statistics, where the energies are given in
Eqs.~(\ref{eq:en-lowest}), (\ref{eq:eigen-energy}). We note that the
energy for $n>0$ is an even function of $p_{z}$ while that for the
lowest level is not if $\mu_{5}$ is non-zero. 

As a final remark we would like to point out that, if the magnetic field is strong enough so that only the
lowest Landau level is occupied, due to the large energy gap between
the lowest and the higher Landau levels the transport coefficients
have very special properties \cite{Hattori:2016cnt,Hattori:2016lqx}.
Nevertheless, one can construct an effective theory, and even fluid dynamics, for a system
where only the lowest Landau level is occupied \cite{Geracie:2014zha}.

\section{Wigner Functions}

\label{sec:Wigner-Functions}
The gauge-invariant Wigner function for fermions 
is defined by \cite{Elze:1986hq,Elze:1986qd,Vasak:1987um,Elze:1989un}
\begin{eqnarray}
W_{\alpha\beta}(X,P) & = & \int\frac{d^{4}X^{\prime}}{(2\pi)^{4}}\exp(-ip_{\mu}x^{\prime\mu})
\left\langle \bar{\psi}_{\beta}\left(X+\frac{1}{2}X^{\prime}\right)U
\left(A,X+\frac{1}{2}X^{\prime},X-\frac{1}{2}X^{\prime}\right)\psi_{\alpha}\left(X-\frac{1}{2}X^{\prime}\right)\right\rangle ,
\end{eqnarray}
where $U\left(A,X+\frac{1}{2}X^{\prime},X-\frac{1}{2}X^{\prime}\right)$ 
is the gauge link between $X-\frac{1}{2}X^{\prime}$ and $X+\frac{1}{2}X^{\prime}$.
Since we consider a constant and homogeneous external magnetic field along the $z$
direction, for which the electromagnetic gauge potential can be chosen
as $A^{\mu}(X)=(0,-By,0,0)$, cf.\ Eq.~(\ref{eq:vector-pot}), the
gauge link is just a phase, $U\left(A,X+\frac{1}{2}X^{\prime},X-\frac{1}{2}X^{\prime}\right)=\exp\left(-ieByx^{\prime}\right)$.
Thus the Wigner function is given by 
\begin{eqnarray}
W(X,P) & = & \int\frac{d^{4}X^{\prime}}{(2\pi)^{4}}\exp\left(-ip_{\mu}x^{\prime\mu}-ieByx^{\prime}\right)
\left\langle \bar{\psi}\left(X+\frac{1}{2}X^{\prime}\right)\otimes\psi\left(X-\frac{1}{2}X^{\prime}\right)\right\rangle .
\label{eq:Wigner}
\end{eqnarray}
The Wigner function can be decomposed in terms of the 16 independent
generators of the Clifford algebra \cite{Vasak:1987um}, 
\begin{eqnarray}
W(X,P) & = & \frac{1}{4}\left(\mathcal{F}+i\gamma^{5}\mathcal{P}+\gamma^{\mu}\mathcal{V}_{\mu}+
\gamma^{5}\gamma^{\mu}\mathcal{A}_{\mu}+\frac{1}{2}\sigma^{\mu\nu}\mathcal{S}_{\mu\nu}\right),
\label{eq:decomp-W}
\end{eqnarray}
where the coefficients $\mathcal{F}$, $\mathcal{P}$, $\mathcal{V}_{\mu}$,
$\mathcal{A}_{\mu}$, and $\mathcal{S}_{\mu\nu}$ are the scalar, pseudo-scalar,
vector, axial-vector, and tensor components of the Wigner function,
respectively. The tensor component is anti-symmetric so we can equivalently
introduce two vector functions 
\begin{eqnarray}
\mathcal{\boldsymbol{T}}=\frac{1}{2}\mathbf{e}_{i}(\mathcal{S}^{0i}-\mathcal{S}^{i0}) 
& , & \mathcal{\boldsymbol{S}}=\frac{1}{2}\epsilon_{ijk}\mathbf{e}_{i}\mathcal{S}_{jk}.
\end{eqnarray}
The functions $\mathcal{F}$, $\mathcal{P}$, $\mathcal{V}_{\mu}$,
$\mathcal{A}_{\mu}$, $\mathcal{\boldsymbol{T}}$, and $\mathcal{\boldsymbol{S}}$
are called Dirac-Heisenberg-Wigner (DHW) functions. All of them are
real functions over phase space and some of them have an obvious physical
meaning \cite{BialynickiBirula:1991tx}. For example, $\mathcal{V}_{\mu}(X,P)$
is the fermion-current density. 

In order to determine the DHW functions in a constant magnetic field,
we insert the field operator (\ref{eq:field-operator-1}) into
the definition (\ref{eq:Wigner}) of the Wigner function. The only
combinations of creation and annihilation operators which survive
when ensemble-averaging are $a_{p_{x}p_{z}s}^{(n)\dagger}a_{p_{x}p_{z}s}^{(n)}=\hat{n}_{p_{x}p_{z}s}^{(n)}$
and $b_{p_{x}p_{z}s}^{(n)\dagger}b_{p_{x}p_{z}s}^{(n)}=\hat{\bar{n}}_{p_{x}p_{z}s}^{(n)}$.
These have been calculated in the previous section, see Eqs.~(\ref{eq:part-n}), (\ref{eq:ant-part-n}).
Since we assume constant chemical potentials and temperature, the
Wigner function does not depend on space-time,
\begin{eqnarray}
W(P) & = & \sum_{n,s}\left\{ f_{\mathrm{FD}}(E_{p_{z}s}^{(n)}-\mu)\delta(p_{0}+\mu-E_{p_{z}s}^{(n)})
W_{+,s}^{(n)}(\mathbf{p})\right.\nonumber \\
 &  & \left.+[1-f_{\mathrm{FD}}(E_{p_{z},s}^{(n)}+\mu)]\delta(p_{0}+\mu+E_{p_{z},s}^{(n)})
 W_{-,s}^{(n)}(\mathbf{p})\right\} .\label{eq:w-x-p}
\end{eqnarray}
Here the 1 in the square brackets is the vacuum contribution arising
from the anti-commutation relation for $b_{p_{x}p_{z}s}^{(n)\dagger}$,
$b_{p_{x}p_{z}s}^{(n)}$. We will show in Sec.~\ref{sec:Chiral-Effects}
that this vacuum term contributes to the chiral magnetic effect. The
matrix-valued functions $W_{\pm,s}^{(n)}(\mathbf{p})$ denote the
contributions of fermions/anti-fermions in the $n$-th Landau level
with $E_{p_{z}s}^{(n)}$. They are straightforwardly computed as 
\begin{eqnarray}
W_{rs}^{(n)}(\mathbf{p}) & \equiv & \frac{1}{(2\pi)^{3}}\int dy^{\prime}\exp\left(ip_{y}y^{\prime}\right)\xi_{rs}^{(n)\dagger}
\left(p_{x},p_{z},\frac{1}{2}y^{\prime}\right)\gamma^{0}\otimes\xi_{rs}^{(n)}\left(p_{x},p_{z},-\frac{1}{2}y^{\prime}\right),
\label{eq:W-n-r}
\end{eqnarray}
where we used the property 
$\phi_{n}\left(p_{x}-eBy,y-\frac{1}{2}y^{\prime}\right)=\phi_{n}\left(p_{x},-\frac{1}{2}y^{\prime}\right)$
and the fact that the dependence of $\xi_{rs}^{(n)}$ on $p_{x}$
and $y$ only appear in the eigenfunctions $\phi_{n}$ and $\phi_{n-1}$
of the harmonic oscillator, see Eqs.~(\ref{eq:wave}), (\ref{eq:chi-0}), (\ref{eq:chi-1}), and (\ref{eq:xi-rs}). 

The functions $W_{r}^{(0)}(\mathbf{p})$ and $W_{rs}^{(n)}(\mathbf{p})$
are evaluated in Appendix \ref{sec:app1}. The results are given in
Eqs.~(\ref{eq:w0}), (\ref{eq:wn}) and Eqs.~(\ref{eq:w0-result}), (\ref{eq:wn-result}).
We can extract all DHW functions from the Wigner function (\ref{eq:w-x-p})
with $W_{r}^{(0)}(\mathbf{p})$ and $W_{rs}^{(n)}(\mathbf{p})$ given
by Eqs.~(\ref{eq:w0-result}), (\ref{eq:wn-result}). In order to write
these functions in compact form, we divide the 16 DHW functions into
four groups, each group forming a four-dimensional vector,
\begin{align}
\boldsymbol{G}_{1}(P)\equiv\left(\begin{array}{c}
\mathcal{F}(P)\\
\mathcal{\boldsymbol{S}}(P)
\end{array}\right),\;\;\; & \boldsymbol{G}_{2}(P)\equiv\left(\begin{array}{c}
\mathcal{V}_{0}(P)\\
\mathcal{\boldsymbol{A}}(P)
\end{array}\right),\nonumber \\
\boldsymbol{G}_{3}(P)\equiv\left(\begin{array}{c}
\mathcal{A}_{0}(P)\\
\mathcal{\boldsymbol{V}}(P)
\end{array}\right),\;\;\; & \boldsymbol{G}_{4}(P)\equiv\left(\begin{array}{c}
\mathcal{P}(P)\\
\mathcal{\boldsymbol{T}}(P)
\end{array}\right).\label{eq:G-i-DHW}
\end{align}
All of these are functions of four-momentum $P$. In order to separate
the $\mathbf{p}_{T}$ dependence we define four-dimensional vectors for $n\geq 0$
\begin{equation}
\boldsymbol{e}_{1}^{(n)}(p_{T})=\left(\begin{array}{c}
\Lambda_{+}^{(n)}(p_{T})\\
\boldsymbol{0}_{T}\\
\Lambda_{-}^{(n)}(p_{T})
\end{array}\right),\ \boldsymbol{e}_{2}^{(n)}(\mathbf{p})=\left(\begin{array}{c}
p_{z}\Lambda_{-}^{(n)}(p_{T})\\
\frac{2neB}{p_{T}^{2}}\Lambda_{+}^{(n)}(p_{T})\mathbf{p}_{T}\\
p_{z}\Lambda_{+}^{(n)}(p_{T})
\end{array}\right).\label{eq:e-i}
\end{equation}
Here $\mathbf{0}_{T}=(0,0)^{T}$ is a two-dimensional null vector and $\Lambda ^{(n)}_{\pm}$ 
is defined in Eqs. (\ref{lambda_pm1},\ref{lambda_pm2}). Then, the DHW functions read
\begin{eqnarray}
\left(\begin{array}{c}
\boldsymbol{G}_{1}(P)\\
\boldsymbol{G}_{2}(P)
\end{array}\right) & = & \left[\sum_{n=0}^{\infty}V_{n}(p_{0},p_{z})\boldsymbol{e}_{1}^{(n)}(p_{T})
+\sum_{n=1}^{\infty}\frac{1}{\sqrt{p_{z}^{2}+2neB}}A_{n}(p_{0},p_{z})\boldsymbol{e}_{2}^{(n)}(\mathbf{p})\right]
\left(\begin{array}{c}
m\\
p_{0}+\mu
\end{array}\right),\nonumber \\
\boldsymbol{G}_{3}(P) & = & (p_{z}-\mu_{5})V_{0}(p_{0},p_{z})\boldsymbol{e}_{1}^{(0)}(p_{T})\nonumber \\
 &  & +\sum_{n=1}^{\infty}\left[\sqrt{p_{z}^{2}+2neB}A_{n}(p_{0},p_{z})-\mu_{5}V_{n}(p_{0},p_{z})\right]
 \boldsymbol{e}_{1}^{(n)}(p_{T})\nonumber \\
 &  & +\sum_{n=1}^{\infty}\left[V_{n}(p_{0},p_{z})-\frac{\mu_{5}}{\sqrt{p_{z}^{2}+2neB}}A_{n}(p_{0},p_{z})\right]
 \boldsymbol{e}_{2}^{(n)}(\mathbf{p}),\nonumber \\
\boldsymbol{G}_{4}(P) & = & 0.\label{eq:G-i}
\end{eqnarray}
Here $V_{n}$ and $A_{n}$ for $n>0$ are given by 
\begin{eqnarray}
V_{n}(p_{0},p_{z}) & = & \frac{2}{(2\pi)^{3}}\sum_{s}\delta\left\{ (p_{0}+\mu)^{2}-[E_{p_{z}s}^{(n)}]^{2}\right\} 
\left\{ \theta(p_{0}+\mu)f_{\mathrm{FD}}(p_{0})+\theta(-p_{0}-\mu)\left[f_{\mathrm{FD}}(-p_{0})-1\right]\right\} ,\nonumber \\
A_{n}(p_{0},p_{z}) & = & \frac{2}{(2\pi)^{3}}\sum_{s}s\delta\left\{ (p_{0}+\mu)^{2}-[E_{p_{z}s}^{(n)}]^{2}\right\} 
\left\{ \theta(p_{0}+\mu)f_{\mathrm{FD}}(p_{0})+\theta(-p_{0}-\mu)\left[f_{\mathrm{FD}}(-p_{0})-1\right]\right\} .\label{eq:A-n}
\end{eqnarray}
The lowest Landau level does not depend on helicity $s$, thus 
\begin{eqnarray}
V_{0}(p_{0},p_{z}) & = & \frac{2}{(2\pi)^{3}}\delta\left\{ (p_{0}+\mu)^{2}-[E_{p_{z}}^{(0)}]^{2}\right\} 
\left\{ \theta(p_{0}+\mu)f_{\mathrm{FD}}(p_{0})+\theta(-p_{0}-\mu)\left[f_{\mathrm{FD}}(-p_{0})-1\right]\right\} .\label{eq:V-0}
\end{eqnarray}

In this paper we choose the Landau gauge when solving the Dirac equation.
If we choose a different gauge, the single-particle wave functions
will be different, but the gauge link in the definition of the Wigner
function will change at the same time, so that the latter is gauge
invariant.

\section{Fermion-number current and chiral-charge current}

\label{sec:Chiral-Effects}
In this section, we will derive the fermion-number current 
$j^{\mu}$ and chiral-charge current $j_{5}^{\mu}$ 
from the DHW functions $\mathcal{V}^{\mu}$ and $\mathcal{A}^{\mu}$,
respectively, by integrating over four-momentum $P$, 
\begin{eqnarray}
j^{\mu} & = & \int d^{4}P\mathcal{V}^{\mu}(P),\nonumber \\
j_{5}^{\mu} & = & \int d^{4}P\mathcal{A}^{\mu}(P).
\end{eqnarray}
The analytical formulas for these DHW functions are given by Eq.~(\ref{eq:G-i}).
The terms corresponding to $\boldsymbol{e}_{2}^{(n)}(\mathbf{p})$
vanish when integrating over $P$ because they are odd under $\mathbf{p}\rightarrow-\mathbf{p}$.
This means that the $x$ and $y$ components of the functions $\boldsymbol{G}_{i}(P)$,
$i=1,2,3,4$, vanish. Thus, also the $x$ and $y$ components of the
fermion-number and chiral-charge currents are zero, which is a consequence
of fact that the motion of the particles is confined to Landau orbits
in the transverse plane.

\subsection{Fermion-number and current density}

The $t$ and $z$ component of the fermion-number current, which denote
the fermion-number density and the current density pointing along the magnetic field, respectively,
are non-zero. From Eqs.~(\ref{eq:G-i-DHW}), (\ref{eq:e-i}), and (\ref{eq:G-i})
we get
\begin{eqnarray}
\rho & = & 2\pi eB\int dp_{0}dp_{z}\sum_{n=0}^{\infty}(p_{0}+\mu)V_{n}(p_{0},p_{z}),\nonumber \\
j_{z} & = & 2\pi eB\int dp_{0}dp_{z}(p_{z}-\mu_{5})V_{0}(p_{0},p_{z}),\label{eq:j0-jz}
\end{eqnarray}
where the functions $V_{n}(p_{0},p_{z})$ are given in Eqs.~(\ref{eq:A-n}), (\ref{eq:V-0})
and we have used Eq.~(\ref{eq:Lambda-pm}). Note that only the lowest
Landau level contributes to $j_{z}$ \cite{Fukushima:2008xe}. In
order to perform the $p_{0}$ integration we use the following properties
of the Dirac delta function,
\begin{eqnarray}
\delta\left\{ (p_{0}+\mu)^{2}-[E_{p_{z}s}^{(n)}]^{2}\right\} \theta\left[r(p_{0}+\mu)\right] 
& = & \frac{1}{2E_{p_{z}s}^{(n)}}\delta(p_{0}+\mu-rE_{p_{z}s}^{(n)}),\nonumber \\
(p_{0}+\mu)\delta\left\{ (p_{0}+\mu)^{2}-[E_{p_{z}s}^{(n)}]^{2}\right\} \theta\left[r(p_{0}+\mu)\right] 
& = & \frac{r}{2}\delta(p_{0}+\mu-rE_{p_{z}s}^{(n)}).\label{eq:delta}
\end{eqnarray}
Here $r=\pm$ for fermions or anti-fermions. Then the fermion-number and current
densities can be expressed as
\begin{eqnarray}
\rho & = & \frac{eB}{(2\pi)^{2}}\sum_{n,s}\int dp_{z}\left[f_{\mathrm{FD}}(E_{p_{z}s}^{(n)}-\mu)
-f_{\mathrm{FD}}(E_{p_{z}s}^{(n)}+\mu)+1\right],\nonumber \\
j_{z} & = & \frac{eB}{(2\pi)^{2}}\int dp_{z}\frac{p_{z}-\mu_{5}}{E_{p_{z}}^{(0)}}\left[f_{\mathrm{FD}}(E_{p_{z}}^{(0)}-\mu)
+f_{\mathrm{FD}}(E_{p_{z}}^{(0)}+\mu)-1\right].
\end{eqnarray}

\begin{figure}
\caption{\label{fig:5.1}Fermion-number density as function of the magnetic field.}
\includegraphics[width= 0.45 \textwidth]{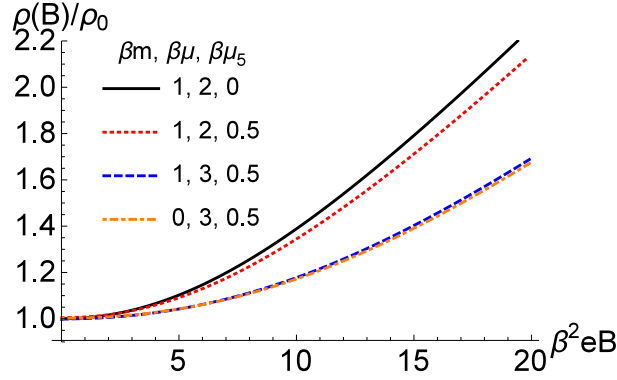}
\end{figure}

In Fig.~\ref{fig:5.1}, we show the ratio of the renormalized fermion-number
density, i.e., the expression without the vacuum term, to the fermion-number
density for $B=0$, 
\begin{eqnarray}
\rho_{0} & = & \frac{1}{(2\pi)^{3}}\int d^{3}\mathbf{p}\sum_{r,s=\pm}rf_{\mathrm{FD}}\left(\sqrt{m^{2}+\mathbf{p}^{2}}
-r\mu+s\mu_{5}\right),\label{eq:n-0}
\end{eqnarray}
as a function of $\beta^{2}eB$. We choose four different configurations:
(i) $\beta m=1$, $\beta\mu=2$, $\beta\mu_{5}=0$, which represents
a chirally symmetric system, (ii) $\beta m=1$, $\beta\mu=2$, $\beta\mu_{5}=0.5$,
representing a system with an imbalance in the number of right- and
left-handed fermions, (iii) $\beta m=1$, $\beta\mu=3$, $\beta\mu_{5}=0.5$,
representing the same case but at a larger fermion-number chemical
potential. The last one, (iv) $\beta m=0$, $\beta\mu=3$, $\beta\mu_{5}=0.5$,
corresponds to the massless case. In all cases the fermion-number
density increases with $B$ and approaches Eq.~(\ref{eq:n-0}) in
the weak-field limit.

We now turn to the computation of the current density. The integration
can be done analytically by first limiting the integration to the region
$\pm\Lambda$ and then taking the limit $\Lambda\rightarrow+\infty$.
The result is 
\begin{eqnarray}
j_{z} & = & -\frac{eB}{4\pi^{2}\beta}\lim_{\Lambda\rightarrow+\infty}\ln\frac{\left\{ 1
+\exp[-\beta(E_{\Lambda}^{(0)}-\mu)]\right\} \left\{ 1+\exp[-\beta(E_{\Lambda}^{(0)}+\mu)]\right\} }{
\left\{ 1+\exp[-\beta(E_{-\Lambda}^{(0)}-\mu)]\right\} \left\{ 1+\exp[-\beta(E_{-\Lambda}^{(0)}+\mu)]\right\} }
-\frac{eB}{4\pi^{2}}\lim_{\Lambda\rightarrow+\infty}\left(E_{\Lambda}^{(0)}-E_{-\Lambda}^{(0)}\right).
\end{eqnarray}
The first term is zero while a careful calculation of the second term
gives
\begin{eqnarray}
j_{z} & = & \frac{e\mu_{5}}{2\pi^{2}}B.
\end{eqnarray}
We have thus reproduced the previous, well-known result for the CME
\cite{Vilenkin:1980fu,Kharzeev:2007jp,Fukushima:2008xe}.

\subsection{Chiral-charge and current density}

Analogously we can derive the chiral-charge and current
densities from Eqs.~(\ref{eq:G-i-DHW}), (\ref{eq:e-i}), and (\ref{eq:G-i}),
\begin{eqnarray}
\rho_{5} & = & 2\pi eB\int dp_{0}dp_{z}\left\{ (p_{z}-\mu_{5})V_{0}(p_{0},p_{z})
+\sum_{n=1}^{\infty}\left[\sqrt{p_{z}^{2}+2neB}A_{n}(p_{0},p_{z})-\mu_{5}V_{n}(p_{0},p_{z})\right]\right\} ,
 \nonumber \\
j_{5z} & = & 2\pi eB\int dp_{0}dp_{z}(p_{0}+\mu)V_{0}(p_{0},p_{z}).
\end{eqnarray}
Using the property (\ref{eq:delta}) of the delta function it is straightforward
to perform the $p_{0}$ integration, 
\begin{eqnarray}
\rho_{5} & = & \frac{eB}{(2\pi)^{2}}\int dp_{z}\left\{ \frac{p_{z}-\mu_{5}}{E_{p_{z}}^{(0)}}
\left[f_{\mathrm{FD}}(E_{p_{z}}^{(0)}-\mu)+f_{\mathrm{FD}}(E_{p_{z}}^{(0)}+\mu)-1\right]\right.\nonumber \\
 &  & \left.+\sum_{n=1}^{\infty}\sum_{s}s\frac{\sqrt{p_{z}^{2}+2neB}-s\mu_{5}}{E_{p_{z}s}^{(n)}}
 \left[f_{\mathrm{FD}}(E_{p_{z}s}^{(n)}-\mu)+f_{\mathrm{FD}}(E_{p_{z}s}^{(n)}+\mu)-1\right]\right\} ,\nonumber \\
j_{5z} & = & \frac{eB}{(2\pi)^{2}}\int dp_{z}\left[f_{\mathrm{FD}}(E_{p_{z}}^{(0)}-\mu)
-f_{\mathrm{FD}}(E_{p_{z}}^{(0)}+\mu)+1\right].\label{eq:n5-j5z}
\end{eqnarray}

\begin{figure}
\caption{\label{fig:5.3} Chiral-charge density as function of the magnetic field.}
\includegraphics{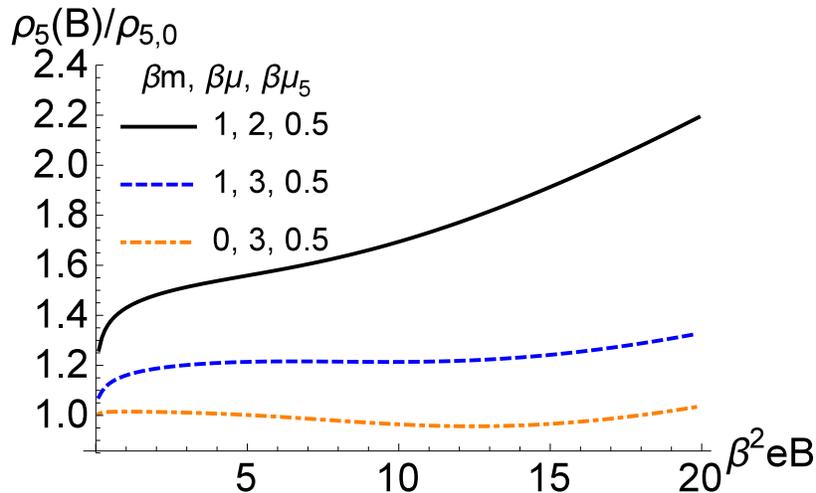}
\end{figure}

Focusing on the chiral-charge density, we numerically perform the
$p_{z}$ integration and compare with the $B=0$ limit, 
\begin{eqnarray}
\rho_{5,0} & = & \frac{1}{(2\pi)^{3}}\int d^{3}\mathbf{p}\sum_{r,s=\pm}rsf_{\mathrm{FD}}
\left(\sqrt{m^{2}+\mathbf{p}^{2}}-r\mu+s\mu_{5}\right).
\end{eqnarray}
The ratio $\rho_{5}(B)/\rho_{5,0}$ is shown in Fig.~\ref{fig:5.3}.
We observe that the ratio is $1$ in the weak-field limit (up to numerical
errors in our algorithm).

We now turn to the chiral-charge current density, cf.\ the last line
of Eq.~(\ref{eq:n5-j5z}). In general, the result cannot be given
in a closed analytic form. As a general remark, however, note that
$j_{5z}$ does not depend on $\mu_{5}$, by virtue of a shift of the
integration variable $p_{z}\rightarrow p_{z}+\mu_{5}$. In the limit
$m\ll T$, we may expand the integrand in a power series in $\beta m$.
The leading term, corresponding to $m=0$, can be analytically calculated.
The first two terms in this expansion read 
\begin{eqnarray}
j_{5z} & = & \frac{eB\mu}{2\pi^{2}}-\frac{eBT(\beta m)^{2}}{(2\pi)^{2}}\int_{0}^{\infty}dp\frac{e^{\beta(p-\mu)}
(e^{2\beta\mu}-1)(e^{2\beta p}-1)}{p[1+e^{\beta(p+\mu)}]^{2}[1+e^{\beta(p-\mu)}]^{2}}+O[(\beta m)^{4}].
\end{eqnarray}
The leading-order term was first calculated by Metlitski and Zhitnitsky
\cite{Metlitski:2005pr} and later reproduced by many groups in different
approaches \cite{Gao:2012ix,Kharzeev:2012ph,Kharzeev:2015znc}.

\section{Summary}

We have computed the covariant Wigner function for spin-1/2 fermions in
an arbitrarily strong magnetic field $\mathbf{B}$ by exactly solving
the Dirac equation at non-zero fermion-number and chiral-charge densities
(or equivalently non-zero chemical potentials $\mu$ and $\mu_{5}$
for the fermion number and chiral charge, respectively). The Landau
energy levels and the corresponding orthonormal eigenfunctions were obtained.
With these orthonormal eigenfunctions we have constructed the fermion field operators
in canonical quantization and, consequently, the Wigner function
operator. The Wigner function was then obtained
by taking the ensemble average of the Wigner function operator in
global thermodynamical equilibrium, i.e., at constant temperature
$T$ and non-zero $\mu$ and $\mu_{5}$. By extracting the vector
and axial-vector components of the Wigner function and carrying out
four-momentum integrals, we obtain the fermion-number and chiral-charge
currents, which agree with the standard results for the CME and CSE, respectively, in an arbitrarily
strong magnetic field. 

\section*{Acknowledgments} 

QW thanks I. Shovkovy for helpful discussions. 
DHR acknowledges support by the High-End Visiting Expert project GDW20167100136 
of the State Administration of Foreign Experts Affairs (SAFEA) of China 
and by the Deutsche Forschungsgemeinschaft (DFG) through the grant CRC-TR 211 
"Strong-interaction matter under extreme conditions". 
QW is supported in part by the Major State Basic Research
Development Program in China (973 program) under the Grant No.\ 2015CB856902
and 2014CB845402 and by the National Natural Science Foundation of
China (NSFC) under the Grant No.\ 11535012. 
This work was first presented at the \textit{Frankfurt Institute of Advanced Studies International Symposium on Discoveries at the Frontiers of Science} held in memory of Walter Greiner (1935-2016). 
We dedicate this work to Walter Greiner, who was teacher, mentor, and friend of DHR, DV, and QW.

When finalizing this work, we became aware that the authors of \cite{Gorbar:2017awz} 
were performing a related study which reaches similar conclusions as our work.

\appendix

\section{Derivation of \textmd{\textup{\normalsize{}$W_{r}^{(0)}(\mathbf{p})$
and}} \textmd{\textup{\normalsize{}$W_{rs}^{(n)}(\mathbf{p})$}}}

\label{sec:app1}In this appendix, we will give the detailed derivation
of $W_{r}^{(0)}(\mathbf{p})$ and $W_{rs}^{(n)}(\mathbf{p})$. Substituting
$\xi_{r}^{(0)}$ and $\xi_{rs}^{(n)}$ from Eqs.~(\ref{eq:chi-0}), (\ref{eq:xi-rs})
into Eq.~(\ref{eq:W-n-r}), we obtain $W_{r}^{(0)}(\mathbf{p})$ for
the lowest Landau level, 
\begin{eqnarray}
W_{r}^{(0)}(\mathbf{p}) & = & \frac{1}{(2\pi)^{3}}\int dy^{\prime}\exp\left(ip_{y}y^{\prime}\right)\xi_{r}^{(0)}
\left(p_{x},p_{z},-\frac{1}{2}y^{\prime}\right)\xi_{r}^{(0)\dagger}\left(p_{x},p_{z},\frac{1}{2}y^{\prime}\right)
\gamma^{0}\nonumber \\
 & = & \frac{1}{(2\pi)^{3}}\frac{1}{2E_{p_{z}}^{(0)}}\int dy^{\prime}\exp\left(ip_{y}y^{\prime}\right)
 \phi_{0}\left(p_{x},-\frac{1}{2}y^{\prime}\right)\phi_{0}\left(p_{x},\frac{1}{2}y^{\prime}\right)\nonumber \\
 &  & \times\left(\begin{array}{cc}
rm & E_{p_{z}}^{(0)}-r(p_{z}-\mu_{5})\\
{}E_{p_{z}}^{(0)}+r(p_{z}-\mu_{5}) & rm
\end{array}\right)\otimes\left(\begin{array}{cc}
1 & 0\\
0 & 0
\end{array}\right).\label{eq:w0}
\end{eqnarray}
We obtain $W_{rs}^{(n)}(\mathbf{p})$ for the higher Landau levels, 
\begin{eqnarray}
W_{rs}^{(n)}(\mathbf{p}) & = & \frac{1}{(2\pi)^{3}}\int dy^{\prime}\exp\left(ip_{y}y^{\prime}\right)\xi_{rs}^{(n)}
\left(p_{x},p_{z},-\frac{1}{2}y^{\prime}\right)\xi_{rs}^{(n)\dagger}\left(p_{x},p_{z},\frac{1}{2}y^{\prime}\right)
\gamma^{0}\nonumber \\
 & = & \frac{1}{(2\pi)^{3}2E_{p_{z}s}^{(n)}}\left(\begin{array}{cc}
rm & E_{p_{z}s}^{(n)}-rs\left(\sqrt{p_{z}^{2}+2neB}-s\mu_{5}\right)\\
E_{p_{z}s}^{(n)}+rs\left(\sqrt{p_{z}^{2}+2neB}-s\mu_{5}\right) & rm
\end{array}\right)\nonumber \\
 &  & \otimes\left(\begin{array}{cc}
c_{n}^{2}I_{nn} & c_{n}d_{n-1}I_{n,n-1}\\
c_{n}d_{n-1}I_{n-1,n} & d_{n-1}^{2}I_{n-1,n-1}
\end{array}\right),\label{eq:wn}
\end{eqnarray}
where the integrals are defined by 
\begin{eqnarray}
I_{ij} & = & \int dy^{\prime}\exp\left(ip_{y}y^{\prime}\right)\phi_{i}\left(p_{x},-\frac{1}{2}y^{\prime}\right)
\phi_{j}\left(p_{x},\frac{1}{2}y^{\prime}\right)\label{eq:integral}
\end{eqnarray}
for $i,j=n$ or $n-1$. The coefficients are evaluated from Eqs.~(\ref{eq:chi-1}), (\ref{eq:cn-dn-1})
as 
\begin{eqnarray}
c_{n}^{2} & = & \frac{1}{2}\left(1+\frac{sp_{z}}{\sqrt{p_{z}^{2}+2neB}}\right),\nonumber \\
c_{n}d_{n-1} & = & \frac{s}{2}\frac{\sqrt{2neB}}{\sqrt{p_{z}^{2}+2neB}},\nonumber \\
d_{n-1}^{2} & = & \frac{1}{2}\left(1-\frac{sp_{z}}{\sqrt{p_{z}^{2}+2neB}}\right).
\end{eqnarray}
For $n>0$, the integrals in Eq.~(\ref{eq:integral}) can be computed
analytically as 
\begin{eqnarray}
\frac{1}{2}\left(I_{n,n}\pm I_{n-1,n-1}\right) & = & \Lambda_{\pm}^{(n)}(p_{T}),\nonumber \\
\frac{1}{2}\left(I_{n,n-1}+I_{n-1,n}\right) & = & \frac{p_{x}\sqrt{2neB}}{p_{T}^{2}}\Lambda_{+}^{(n)}(p_{T}),\nonumber \\
\frac{1}{2}\left(I_{n,n-1}-I_{n-1,n}\right) & = & \frac{ip_{y}\sqrt{2neB}}{p_{T}^{2}}\Lambda_{+}^{(n)}(p_{T}),
\end{eqnarray}
where $p_{T}=\sqrt{p_{x}^{2}+p_{y}^{2}}$ is the modulus of the transverse
momentum and $\Lambda_{\pm}^{(n)}(p_{T})$ ($n>0$) are defined as
\begin{eqnarray}
\Lambda_{\pm}^{(n)}(p_{T}) & = & (-1)^{n}\left[L_{n}\left(\frac{2p_{T}^{2}}{eB}\right)\mp L_{n-1}
\left(\frac{2p_{T}^{2}}{eB}\right)\right]\exp\left(-\frac{p_{T}^{2}}{eB}\right),
\label{lambda_pm1}
\end{eqnarray}
where $L_{n}(x)$ are the Laguerre polynominals with $L_{-1}(x)=0$.
For the lowest Landau level, $n=0$, we have 
\begin{equation}
I_{00}=\Lambda^{(0)}_\pm (p_{T})=\Lambda^{(0)} (p_{T})=2\exp\left(-\frac{p_{T}^{2}}{eB}\right).
\label{lambda_pm2}
\end{equation}
One can check that when integrating over $\mathbf{p}_{T}=(p_{x},p_{y})^T$,
the functions $\Lambda_{+}^{(n)}(p_{T})$ ($n>0$) and $\Lambda^{(0)}(p_{T})$
will give the density of states and $\Lambda_{-}^{(n)}(p_{T})$ ($n>0$)
will give zero, 
\begin{eqnarray}
\int\frac{d^{2}\mathbf{p}_{T}}{(2\pi)^{2}}\Lambda_{+}^{(n)}(p_{T}) 
& = & \int\frac{d^{2}\mathbf{p}_{T}}{(2\pi)^{2}}\Lambda^{(0)}(p_{T})=\frac{eB}{2\pi},\nonumber \\
\int\frac{d^{2}\mathbf{p}_{T}}{(2\pi)^{2}}\Lambda_{-}^{(n)}(p_{T}) & = & 0.\label{eq:Lambda-pm}
\end{eqnarray}

We can expand Eq.~(\ref{eq:w0}) as in Eq.~(\ref{eq:decomp-W}) and
obtain 
\begin{eqnarray}
W_{r}^{(0)}(\mathbf{p}) & = & \frac{r}{4(2\pi)^{3}E_{p_{z}}^{(0)}}\Lambda^{(0)}(p_{T})
\left[m(1+\sigma^{12})+rE_{p_{z}}^{(0)}(\gamma^{0}-\gamma^{5}\gamma^{3})
 -(p_{z}-\mu_{5})(\gamma^{3}-\gamma^{5}\gamma^{0})\right].\label{eq:w0-result}
\end{eqnarray}
Similarly we can also expand Eq.~(\ref{eq:wn}) as 
\begin{eqnarray}
W_{rs}^{(n)}(\mathbf{p}) & = & r\frac{1}{4(2\pi)^{3}E_{p_{z}s}^{(n)}}\left\{ \left[\Lambda_{+}^{(n)}(p_{T})
+s\frac{p_{z}}{\sqrt{p_{z}^{2}+2neB}}\Lambda_{-}^{(n)}(p_{T})\right]
\left[m+rE_{p_{z}s}^{(n)}\gamma^{0}+\left(s\sqrt{p_{z}^{2}+2neB}-\mu_{5}\right)
 \gamma^{5}\gamma^{0}\right] \right.\nonumber \\
 &  & \hspace*{1.9cm}-\left[\Lambda_{-}^{(n)}(p_{T})+s\frac{p_{z}}{\sqrt{p_{z}^{2}+2neB}}\Lambda_{+}^{(n)}(p_{T})\right]
 \left[\left(s\sqrt{p_{z}^{2}+2neB}-\mu_{5}\right)\gamma^{3}+rE_{p_{z}s}^{(n)}\gamma^{5}\gamma^{3}
 -m\sigma^{12}\right]\nonumber \\
 &  & \hspace*{1.9cm}-\frac{2neB}{p_{T}^{2}\sqrt{p_{z}^{2}+2neB}}\Lambda_{+}^{(n)}(p_{T})\left[\left(\sqrt{p_{z}^{2}+2neB}
 -s\mu_{5}\right)\left(p_{x}\gamma^{1}+p_{y}\gamma^{2}\right)\right.\nonumber \\
 &  & \hspace*{6cm}\left.\left.+rsE_{p_{z}s}^{(n)}\left(p_{x}\gamma^{5}\gamma^{1}+p_{y}\gamma^{5}\gamma^{2}\right)
 -sm\left(p_{x}\sigma^{23}-p_{y}\sigma^{13}\right)\right]\right\} .\label{eq:wn-result}
\end{eqnarray}
The global factors $r/2E_{p_{z}}^{(0)}$ and $r/2E_{p_{z}s}^{(n)}$
in front of the brackets will be used later to write the delta functions
in Eq.~(\ref{eq:w-x-p}) in a covariant form. 

\bibliographystyle{apsrev}
\bibliography{ref}

\end{document}